\documentclass[conference]{IEEEtran}
\IEEEoverridecommandlockouts
\usepackage{cite}
\usepackage{amsmath,amssymb,amsfonts}
\usepackage{algorithmic}
\usepackage{graphicx}
\usepackage{textcomp}
\usepackage{xcolor}
\usepackage{booktabs}
\usepackage{url}
\usepackage[top=0.75in, left=0.63in, right=0.63in, bottom=1.04in]{geometry}
\def\BibTeX{{\rm B\kern-.05em{\sc i\kern-.025em b}\kern-.08em
    T\kern-.1667em\lower.7ex\hbox{E}\kern-.125emX}}
\begin{document}

\title{DisfluencySpeech -- Single-Speaker Conversational Speech Dataset with Paralanguage\\
\thanks{This project has received funding from SUTD Kickstarter Initiative no. SKI 2021\_04\_06.}
}

\author{\IEEEauthorblockN{Kyra Wang}
\IEEEauthorblockA{\textit{Information Systems Technology and Design} \\
\textit{Singapore University of Technology and Design}\\
Singapore, Singapore \\
kyra\_wang@mymail.sutd.edu.sg}
\and
\IEEEauthorblockN{Dorien Herremans}
\IEEEauthorblockA{\textit{Information Systems Technology and Design} \\
\textit{Singapore University of Technology and Design}\\
Singapore, Singapore \\
dorien\_herremans@sutd.edu.sg}
}

\maketitle

\begin{abstract}
Laughing, sighing, stuttering, and other forms of paralanguage do not contribute any direct lexical meaning to speech, but they provide crucial propositional context that aids semantic and pragmatic processes such as irony. It is thus important for artificial social agents to both understand and be able to generate speech with semantically-important paralanguage. Most speech datasets do not include transcribed non-lexical speech sounds and disfluencies, while those that do are typically multi-speaker datasets where each speaker provides relatively little audio. This makes it challenging to train conversational Text-to-Speech (TTS) synthesis models that include such paralinguistic components.

We thus present DisfluencySpeech, a studio-quality labeled English speech dataset with paralanguage. A single speaker recreates nearly 10 hours of expressive utterances from the Switchboard-1 Telephone Speech Corpus (Switchboard), simulating realistic informal conversations. To aid the development of a TTS model that is able to predictively synthesise paralanguage from text without such components, we provide three different transcripts at different levels of information removal (removal of non-speech events, removal of non-sentence elements, and removal of false starts), as well as benchmark TTS models trained on each of these levels.
\end{abstract}

\begin{IEEEkeywords}
artificial intelligence, speech synthesis, dataset, text-to-speech, disfluency
\end{IEEEkeywords}

\section{Introduction}
In informal spoken English, while the content of what is spoken is important for deriving meaning, how that content is said is just as important. While non-speech sounds (such as laughter and sighing) and disfluencies (such as filled pauses like ``um'' and ``uh'') do not directly contribute lexical information, it has been argued that such non-lexical components of conversational speech, or paralanguage, can drastically influence the perception of what is said, altering the semantic context of the lexical content \cite{ginzburgLaughterLanguage2020}.

Examples of the semantic and pragmatic significance of paralanguage in speech can be found in a multitude of examples: conversational grunts have a wide variety of pragmatic functions ranging from indicating interest in an interaction to controlling the flow of turn-taking \cite{wardNonlexicalConversationalSounds2006}; disfluent speech can affect a listener's interpretation of the deceptiveness of a message \cite{loyEffectsDisfluencyOnline2017}; even from a young age toddlers use speech disfluencies for realtime comprehension of speaker referential intention \cite{kiddToddlersUseSpeech2011}.

Systems capable of synthesising speech with semantically-appropriate paralinguistic components are thus very useful in the many areas that can benefit from an additional dimension of information representation, such as the design of the personality of synthesised voices for human computer interaction \cite{leeDesigningSocialPresence2003}.

In regard to modern neural speech generation, of the many forms of paralanguage, prosodic speech is probably the most well investigated field \cite{tanSurveyNeuralSpeech2021}. Synthesis of disfluent speech and non-speech sounds, on the other hand, are relatively unexplored areas. \cite{schullerComputationalParalinguisticsEmotion2013} argues that this is due to the complexity of the task of modelling these paralinguistic components of speech explicitly.

Computational modelling of these components is hampered in particular by the lack of existing speech datasets that include them. Most popular speech datasets are either multi-speaker datasets that do not include enough data per speaker (such as the Switchboard-1 Telephone Speech Corpus (Switchboard) \cite{godfreyjohnj.Switchboard1Release1993}), or single-speaker datasets that do not include non-lexical speech sounds and disfluencies (such as LJSpeech \cite{itoLJSpeechDataset2017}). This makes it challenging to train conversational TTS synthesis models that include such paralinguistic components.

The closest work to providing a dataset that includes disfluencies with considerable audio data from each speaker is DailyTalk \cite{leeDailyTalkSpokenDialogue2022}. However, while it contains sample dialogues from the DailyDialog dataset \cite{liDailyDialogManuallyLabelled2017} to replicate natural speech, it does not include disfluencies other than filled pauses such as ``um'' and ``uh'', and does not include non-lexical speech sounds such as laughter and sighing.

To that end, in this paper, we present DisfluencySpeech, a single-speaker, studio-quality, fully-labeled, English speech dataset. The dataset consists of nearly 10 hours of utterances derived from the Switchboard corpus. Unlike the DailyTalk dataset, our dataset includes detailed disfluency and non-lexical speech sound annotations, distinguishing between different types such as filled pauses and laughter. We also provide three different transcripts at different levels of information removal (removal of non-speech events, removal of non-sentence elements, and removal of false starts), to aid the development of a TTS model that is able to predictively synthesise semantically-meaningful paralanguage from text without such components \footnote{\url{https://huggingface.co/datasets/amaai-lab/DisfluencySpeech}}.

Additionally, to help users of this dataset speed up their analysis, we provide weights for benchmark Transformer \cite{liNeuralSpeechSynthesis2019} models trained on each transcript, a HifiGAN vocoder \cite{kongHiFiGANGenerativeAdversarial2020} fine-tuned on the dataset, and Montreal Forced Aligner (MFA) \cite{mcauliffeMontrealForcedAligner2017} resources adapted to the dataset.

\section{DisfluencySpeech dataset}

The construction of the dataset comprised of three steps: generating the transcript, recording a single speaker reading the transcript in a conversational manner, and processing the recorded clips to make them ideal for training TTS models on.

\subsection{Transcript generation}

The transcripts for DisfluencySpeech were derived from the Switchboard Dialog Act Corpus (SwDA) \cite{Jurafsky-etal:1997}, which extends the Switchboard-1 Telephone Speech Corpus, Release 2 \cite{godfreyjohnj.Switchboard1Release1993}. Subutterances were joined together to form full utterances, removing interleaving interruptions from the other speaker. Only utterances with 15 to 35 words were included in the final transcript, to ensure that each utterance was long enough to have meaningful contextual semantic information, but short enough to be easily loaded into memory for training a TTS. These transcripts were manually created and checked for accuracy to each associated audio file.

Five types of non-sentence elements based on \cite{meteerDysfluencyAnnotationStylebook} are annotated in the Switchboard transcript: filled pauses \{F ...\} (e.g. ``uh'', ``um''), explicit editing terms \{E ...\} (e.g. ``I mean'', ``sorry''), discourse markers \{D ...\} (e.g. ``you know'', ``well''), coordinating conjunctions \{C ...\} (e.g. ``and'', ``but''), and asides \{A ...\} (comments that interrupt fluent flow). Additionally, restarts [... + ...] and non-speech sounds <...> are annotated as well.

Three different transcripts are provided with DisfluencySpeech, each at differing levels of information removal:

\begin{itemize}
\item Transcript A contains all textual content recorded, including non-sentence elements and restarts. Only non-speech events such as laughter and sighs are removed from transcript.
\item Transcript B is transcript A but with filled pauses, explicit editing terms, and discourse markers removed. Coordinating conjunctions and asides are left in, as they are non-sentence elements as well, they are often used to convey meaning.
\item Transcript C is transcript B but with false starts removed. This is the most minimal transcript.
\end{itemize}

\begin{figure*}[h!]
  \centering
  \includegraphics[width=\linewidth]{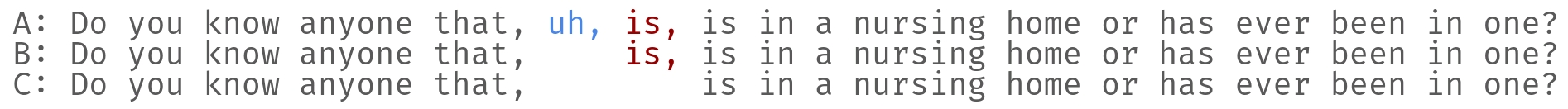}
  \caption{Example of the three different ways that the same clip is transcribed in the transcripts. Blue represents a filled pause, and red a false start.}
  \label{fig:transcripts}
\end{figure*}

An example of the three different ways that the same clip is transcribed in the transcripts can be seen in Figure~\ref{fig:transcripts}.

The reason for providing different levels of transcipts is to explore the possibility of training TTS models that can understand the semantic meaning of the text, and synthesise paralanguage that is semantically-appropriate. For example, a TTS model trained on transcript A may be able to synthesise a sigh when the text contains a discourse marker; more impressively, a TTS model trained on transcript C may be able to synthesise false starts and stuttering when it predicts that the speaker would be nervous or stressed from the text.

These transcripts are provided in the same format as LJ Speech \cite{itoLJSpeechDataset2017} such that they are compatible with any pipelines that use LJ Speech. Each transcript is provided as a .csv file, with one record per line, delimited by the pip character (0x7c). The fields are:

\begin{enumerate}
\item ID: the name of the corresponding .wav file.
\item Transcription: words spoken by the reader (UTF-8). 
\item Normalized Transcription: transcription with numbers, ordinals, and monetary units expanded into full words (UTF-8). This field is redundant as the original Switchboard transcript is already normalized, but is kept for compatibility with existing pipelines.
\end{enumerate}

A transcript of the original joined subutterances extracted from SwDA is also provided, retaining all non-speech event and disfluency annotations, allowing users of the dataset to generate their own transcripts or perform semantic analysis on the dataset.

\subsection{Recording}

Recording was conducted in a acoustically-treated professional studio, using a Samson Q2U dynamic microphone. A pop filter was used to eliminate popping sounds from plosives during recorded speech. Elements of the recording environment were controlled between recording sessions, such as the layout of the furniture in the studio and the distance of the speaker from the microphone, to ensure a consistent sound profile across recorded clips.

A single 28 year old Singaporean speaker with English as her first language is chosen to read the original transcript generated with all Switchboard annotations including non-speech sounds, simulating a conversation with herself.

The format of the audio files, once again, are the same as in LJ Speech \cite{itoLJSpeechDataset2017} to allow compatibility with any pipelines for LJ Speech. Each audio file is a single-channel 16-bit PCM WAV with a sample rate of 22,050 hertz.

\subsection{Post-processing}

A Montreal Forced Aligner \cite{mcauliffeMontrealForcedAligner2017} acoustic model was adapted from the English (US) ARPA acoustic model v2.0.0 \cite{mfa_english_us_arpa_acoustic_2022} onto the dataset, and a corresponding grapheme-to-phoneme (G2P) dictionary was generated using the English (US) ARPA G2P model \cite{mfa_english_us_arpa_g2p_2022}. Users of the dataset who require forced alignment (e.g. providing alignments for non-autoregressive TTS models like FastSpeech~2 \cite{renFastSpeechFastHighQuality2022}) may use these resources. Table~\ref{tab:statistics} shows statistics of the final DisfluencySpeech dataset. 

\begin{table}[htbp]
  \caption{Detailed statistics of DisfluencySpeech}
  \label{tab:statistics}
  \begin{center}
  \begin{tabular}{lc}
    \toprule
    \textbf{Feature}                    & \textbf{Total}      \\
    \midrule
    \# clips                            & $5,000$              \\
    \# words                            & $113,367$            \\
    Total duration (s)                  & $34,146$             \\
    Mean duration/clip (s)              & $6.829$             \\
    Mean \# phone/clip                  & $71.1034$           \\
    \# distinct words                   & $10,703$             \\
    \# clips with filled pause          & $22,74$              \\
    \# clips with explicit editing term & $321$               \\
    \# clips with discourse marker      & $2,000$              \\
    \# clips with restarts              & $2,635$              \\
    \bottomrule
  \end{tabular}
  \end{center}
\end{table}

\section{Benchmark Models}

To assess the usability of the DisfluencySpeech dataset for training TTS models, we trained simple benchmark models for each of the transcripts in the dataset. The benchmark model architecture is the same for each transcript we trained on, and consists of a Transformer \cite{liNeuralSpeechSynthesis2019} autoregressive TTS model implemented using the fairseq $S^2$ speech synthesis toolkit \cite{wangFairseqScalableIntegrable2021}. The 5,000 clips in the dataset were split into a 90\%-5\%-5\% train-validation-test set. During training, each benchmark model took its respective transcript (i.e. A, B, and C) as input, and used the same DisfluencySpeech dataset audio as targets. The hyperparameters used were the same as the ones found on the fairseq $S^2$ LJ Speech Transformer model\footnote{\url{https://github.com/facebookresearch/fairseq/blob/main/examples/speech_synthesis/docs/ljspeech_example.md}}. This is to ensure a fair comparison between DisfluencySpeech and an established dataset, LJSpeech. The training of each model was conducted on a single Tesla V100-DGXS for 30,000 steps.

Automatic evaluation of the benchmark models was performed. The mean cepstral distortion (MCD) was computed on Griffin-Lim \cite{griffinSignalEstimationModified1984} vocoded benchmark model output spectrograms against Griffin-Lim vocoded ground truth spectrograms at a sample rate of 22,050 hertz. In addition, a HiFiGAN vocoder \cite{kongHiFiGANGenerativeAdversarial2020} was used to generate waveforms at a sample rate of 16,000 hertz for character error rate (CER) evaluation using the Wav2Vec 2.0 Large (LV-60 + CV + SWBD + FSH) automatic speech recognition (ASR) model finetuned on 300 hours of the Switchboard dataset \cite{baevskiWav2vecFrameworkSelfSupervised2020}. The HifiGAN vocoder model was finetuned on the dataset using NVIDIA's NeMo toolkit \cite{kuchaievNeMoToolkitBuilding2019}. 

\section{Objective Evaluations}

Table~\ref{tab:results} shows the results of the evaluation. In addition to evaluating the benchmark models, we added metrics calculated on the original datasets themselves and we resynthesize the audio from DisfluencySpeech using HiFiGAN. To calculate the CER between audio and text, we first transcribe the audio using the Wav2Vec 2.0 Large (LV-60 + CV + SWBD + FSH) automatic speech recognition (ASR) model finetuned on 300 hours Switchboard \cite{baevskiWav2vecFrameworkSelfSupervised2020}, and then compare to their respective transcripts. The original DisfluencySpeech audio and resynthesized audio was evaluated against transcript A.

The low CER of the original audio, even when using a pre-trained ASR model not fine-tuned to the speaker's accent, shows that the transcript accurately reflects the recorded speech. The CER is only slightly greater than the CER of LJSpeech, which, unlike DisfluencySpeech, is a speech dataset with no disfluencies and non-speech sounds, elements that greatly increase the chances of error in ASR tasks. Additionally, with the resynthesised audio performing similarly to the original audio in the ASR task, we can conclude that the fine-tuned HiFiGAN model performs well too.

The training of the Transformer model for transcript A was able to converge, and the model can generate speech that is similar to the original audio. The MCD is slightly better than the MCD of the same Transformer model trained on LJSpeech by \cite{wangFairseqScalableIntegrable2021}, indicating that the generated audio is phonetically similar to the original audio. However, the CER is much higher. This is in line with prior research that shows Transformer has trouble generalizing alignments to long utterances \cite{badlaniOneTTSAlignment2022}, which can be seen in the generated audio as frequently missing words, or repeating words too many times.

\begin{table}[htpb]
  \caption{Metrics for the test sets of both datasets, as well as audio generated (from the test set) using the benchmark Transformer (T) model trained on the listed training dataset.}
  \label{tab:results}
  \begin{center}
  \begin{tabular}{llcc}
    \toprule
    \textbf{Dataset}$^{\mathrm{a}}$    & Model     & \textbf{MCD}$^{\mathrm{b}}$ & \textbf{CER}$^{\mathrm{c}}$     \\
    \midrule
    LJSpeech Original      & NA & ---          & $3.30$            \\
    DisfluencySpeech Original & NA        & ---          & $4.04$           \\
    DisfluencySpeech  & Resynthesised    & ---          & $4.26$           \\
    \midrule
    LJSpeech & benchmark (T)      & $3.90$        & $5.20$            \\
    DisfluencySpeech Transcript A & benchmark (T) & $3.68$       & $15.01$          \\
    DisfluencySpeech Transcript B & benchmark (T) & $5.26$       & $60.07$          \\
    DisfluencySpeech Transcript C & benchmark (T) & $4.87$       & $55.66$              \\
    \bottomrule
  \end{tabular}
  \end{center}
  \scriptsize
   $^{\mathrm{a}}$The LJSpeech results are taken from \cite{wangFairseqScalableIntegrable2021}.
    $^{\mathrm{b}}$Mel cepstral distortion (MCD), measured in decibels (dB).
    $^{\mathrm{c}}$Character error rate (CER), measured in percentage.
\end{table}

The brittleness of Transformer's attention mechanism in situations with missing transcript information becomes even more apparent when we see that the training of the Transformer models for transcript B and transcript C both failed to converge. The Transformer model is unable to learn the alignments of the missing textual transcript information, resulting in a high CER and unintelligible generated audio. This is an expected result with these simple benchmark models, and future research could approach this problem with more recent techniques such as enforcing hard monotonic alignments \cite{kim2020glow}.

Despite the problems with alignment, however, transcript A's Transformer model was able to learn to generate non-speech sounds despite the fact that transcript A does not have any non-speech sounds annotated. We notice that sounds like sighs and laughter, while not as prominently featured as in the original audio, are still present in the generated audio. This shows that the Transformer model was able to implicitly learn to generate non-speech sounds as semantically important to the speaker's intent, despite the lack of explicit annotations in the transcript. 

\section{Discussion and Future Research}

It is not the aim of this paper to develop state-of-the-art model, but merely to provide an objective benchmark with open source evaluation framework which will allow researchers to easily work with our dataset. In such model focused work, subjective evaluation such as Mean Opinion Scores would be essential. Given that this is mainly a dataset introduction paper, it falls outside of our scope.

The joining of subutterances from the Switchboard corpus into single utterances removes the context of interleaving speech between speakers in the recorded dialogues, and thus the semantic information in the paralinguistic components of speech loses that dimensionality. This is a limitation of the dataset, and future work should explore the effects of this limitation on the performance of TTS models trained on DisfluencySpeech.

The dataset was built by reading aloud an existing corpus with indications of the disfluencies, which is a form of acting rather than spontaneous speech production. While this has the advantage of creating a single-speaker dataset from many speakers, the lack of spontaneity might affect the performance of TTS models trained on DisfluencySpeech.

In future work, the transcripts of our dataset may also be useful as an auxiliary task for a TTS model. This may teach the model to model disfluencies of speech, without explicitly being fed them as input text. 

\section{Conclusion}

In this paper, we present the DisfluencySpeech dataset, a single-speaker speech dataset with disfluencies and non-speech sounds. Along with the audio, we have also provided manually-checked transcripts at 3 different levels of information removal, intended for the training of TTS systems capable of predictively generating speech with semantically-meaningful paralanguage from text. The dataset is available open source and can be downloaded together with a dataloader online.






\bibliography{IEEEabrv, bibliography}

\end{document}